\documentclass[twocolumn,showpacs,amsmath,prl]{revtex4-1}

\usepackage{color}
\usepackage{graphicx}
\usepackage{epsfig,psfrag}
\usepackage{amsmath}
\usepackage{amssymb}
\usepackage{amsbsy}
\usepackage{amsthm}
\usepackage{amsfonts}
\usepackage{wasysym}
\usepackage{bbm}
\usepackage{tabularx}
\usepackage{euscript}
\usepackage{color}
\usepackage{enumerate}
\usepackage{amsfonts}
\usepackage{exscale}
\usepackage{bbold}
\usepackage{float}

\usepackage[colorlinks,citecolor=blue]{hyperref}

\renewcommand{\v}[1]{{\boldsymbol{#1}}}

\newcommand{\be}{\begin{eqnarray}}
\newcommand{\ee}{\end{eqnarray}}

\newcommand{\Eq}[1]{equation~(\ref{#1})}
\newcommand{\Fig}[1]{Fig.~\ref{#1}}

\newcommand{\nn}{\nonumber\\}

\newcommand{\s}{{\sigma}}
\def\bea{\begin{eqnarray}}
\def\eea{\end{eqnarray}}

\def\avg#1{\left\langle#1\right\rangle}

\def\Eq#1{Eq.~(\ref{#1})}
\def\Fig#1{Fig.~\ref{#1}}

\begin{document}

\title{The thermal Hall conductance of two doped symmetry-breaking topological insulators}
\author{Zi-Xiang Li$^{1,2}$,}
\author{Dung-Hai Lee$^{1,2}$}
\affiliation{
$^1$ Department of Physics, University of California, Berkeley, CA 94720, USA.\\
$^2$ Materials Sciences Division, Lawrence Berkeley National Laboratory, Berkeley, CA 94720, USA.
}
\begin{abstract}
In this paper we study two models of symmetry-breaking topological insulators. They are the variants of the d-density wave Hamiltonian proposed by Chakravarty, Laughlin, Morr and Nyack\cite{Laughlin} to explain the pseudogap of the cuprates.
After doping, both models exhibit an anomalous thermal Hall effect similar to that reported in Ref.\cite{Taillefer-2019}. Moreover, they also possess hole pockets centered along the Brillouin zone diagonals consistent with the Hall coefficient measured in Ref.\cite{Taillefer1}.
\end{abstract}

\maketitle

\section{Introduction}

Symmetry-protected topological states (SPTs) has attracted lots of interests in recent years. These states do not break any Hamiltonian symmetry and are fully gapped in the bulk. In the presence of boundary, SPTs are characterized by gapless boundary modes.  Importantly, as long as the protection symmetry is unbroken the gapless boundary states are protected.\\

In the presence of spontaneous symmetry breaking the symmetry group  of the Hamiltonian is broken down to a subgroup. Due to the protection by this subgroup, symmetry breaking phases can also be divided into different topological classes.  Transitions between topologically inequivalent symmetry-breaking phases are also either first order or continuous quantum phase transitions. Moreover, the interface between different topological phases must also harbor gapless modes. All of these features are the same as SPTs.\\

In the rest of the paper we consider two models of symmetry-breaking topological insulators. The first model is a gapped version of the Hamiltonian introduced in Ref.\cite{Laughlin}, the second model is introduced in Ref.\cite{Hsu}.
 We will show, after low level of p-type doping doping, these models exhibit hole pockets centered along the Brillouin zone diagonals. Interestingly, they also have the potential of explaining the  unusual thermal Hall effect reported in Ref.\cite{Taillefer-2019}.
\\

In Ref.\cite{Taillefer-2019} it is shown that La$_{2-x}$Sr$_x$CuO$_4$ at $x=0.06$  exhibits an unusual thermal Hall conductivity ($\kappa_{xy}$). This sample is superconducting below $5 K$ and situates close to the boundary of the antiferromagnetic phase. At low temperatures $\kappa_{xy}/T$ is negative and the magnitude rises monotonically with the magnetic field strength. This thermal Hall conductivity is apparently not due to charge carriers. Because according to the Wiedemann-Franz law the latter contribution is negligible. Importantly, this unusal thermal Hall effect is also observed in other  cuprate compounds including La$_{1.6-x}$Nd$_{0.4}$Sr$_x$CuO$_4$,  La$_{1.8-x}$Eu$_{0.2}$Sr$_x$CuO$_4$, and Bi$_2$Sr$_{2-x}$La$_x$CuO$_{6+\delta}$ under restricted conditions. The conditions are (1) the doping concentrations exclude those exhibiting charge order, and (2) the values of temperature and magnetic field are such that superconductivity is suppressed. Most surprisingly, under a 15 T magnetic field the
temperature dependence of $\kappa_{xy}/T$ for the undoped La$_2$CuO$_4$ is very close to that of  La$_{2-x}$Sr$_x$CuO$_4$ at $x=0.06$, suggesting a similar anomalous thermal Hall effect in the parent compound of cuprates !  \\

\section{The model}
\subsection{Model 1: a modified  DDW\cite{Laughlin}}

\begin{figure}[t]
\includegraphics[scale=0.33]{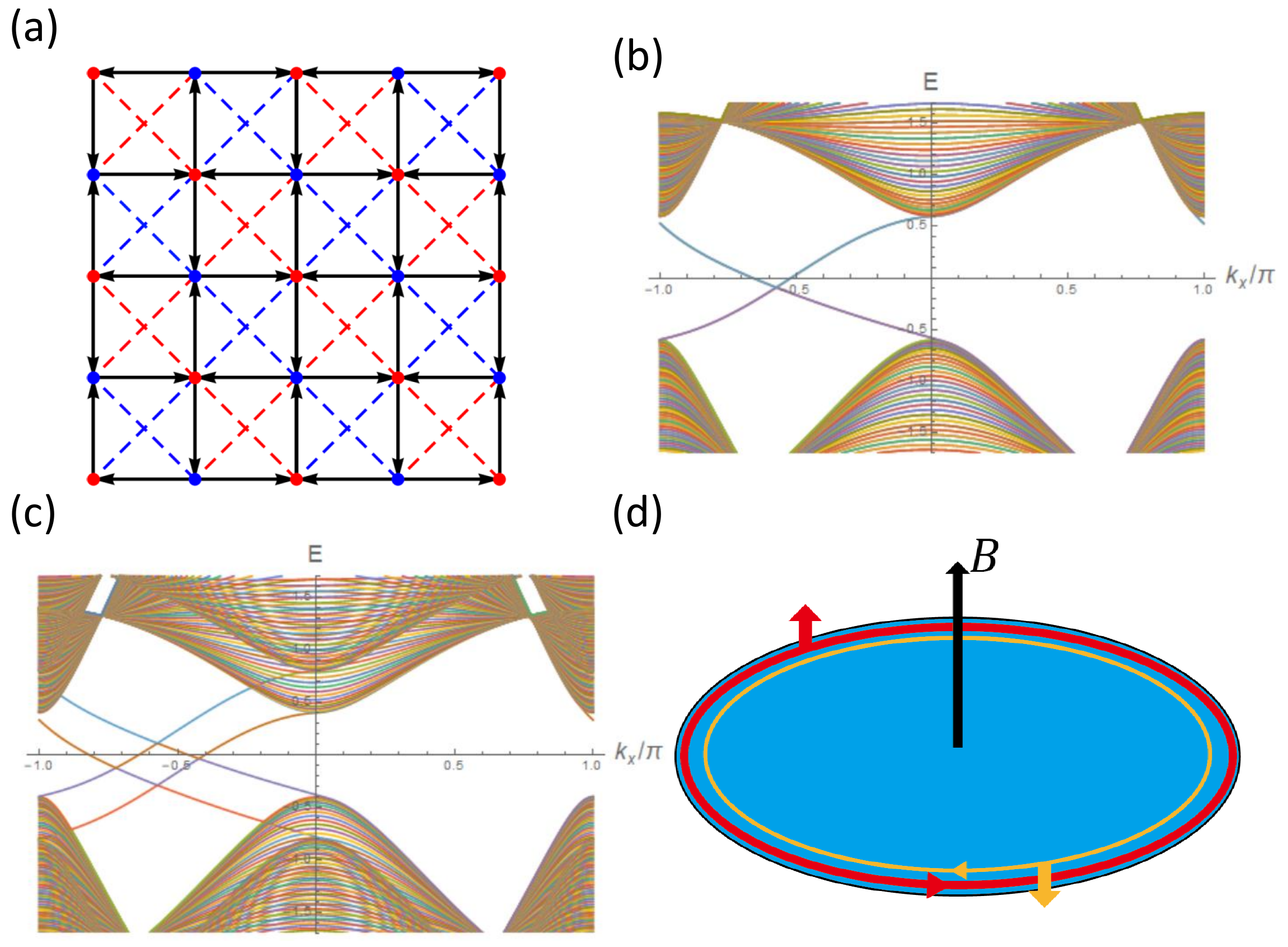}
\caption{ (a) The real-space representation of $H_{\rm s-DDW}$ in \Eq{model}.   The black arrows represent positive imaginary hopping along the designated direction. The hopping amplitude associated with the opposite direction is the complex conjugate. The blue(red) dash line represents positive (negative) real hopping amplitudes.  Note, the translation and rotation symmetries are broken. (b) The projected band structure of \Eq{model} with periodic boundary condition in the $\hat{x}$ and open boundary condition in the $\hat{y}$ directions, respectively. The number of rows in y-direction is $n_y=100$. The parameters used are  $\vec{m}_1 = 0.15 \hat{z}, m_2 = 0.5$. Each of the two in-gap edge branch is two-fold degenerate. This degeneracy is due to the presence of two different edges.  (c) The projected band structure of \Eq{model} in the presence of a magnetic field in the $\hat{z}$ direction. The associated Zeeman energy is set to $0.2$, which causes the edge modes to be Zeeman split. (d) Under an external magnetic field the magnitude of the spin up and spin down edge currents are no longer equal. This results in a  boundary circulating current in the disk geometry. }
\label{fig1}
\end{figure}

The Hamiltonian is given by
 \bea
&&H = H_0 + H_{\textrm{s-DDW}} \nonumber\\
&&H_0 = -t_1 \sum_{\avg{ij}} C^\dagger_{i\alpha} C_{j\alpha} -t_2 \sum_{\avg{\avg{ij}}} C^\dagger_{i\alpha} C_{j\alpha} + h.c. \nonumber\\
&&H_{\textrm{s-DDW}} =\sum_i(-1)^{i_x+i_y}\Big\{i~m_2\Big[C^\dagger_{i\alpha} C_{i+x\alpha} -C^\dagger_{i\alpha}C_{i+y\alpha}\Big]\nn&&+\vec{m}_1\cdot\vec{\s}_{\alpha\beta}\Big[C^\dagger_{i\alpha}C_{i+x+y\beta} - C^\dagger_{i\alpha} C_{i-x+y\beta}\Big] + h.c.\Big\} \nonumber\\
\label{model}
\eea
Here $C_{i,\alpha}$ annihilates a spin $\alpha$ electron on site $i$ of the square lattice, and $\s^{x,y,z}$ are the Pauli matrices. The repeated spin indices  $\alpha,\beta$ imply summation.  $H_0$ describes the dispersion of the Zhang-Rice singlet band. The hopping amplitudes between nearest-neighbor and next-nearest-neighbor sites are $t_1$ and $t_2$, respectively. In the rest of the paper we set $t_1 = 1$ and $t_2 = -0.1$, and denote the values of all other energy parameters in unit of $t_1$. In $H_{\rm s-DDW}$, the term proportional to $im_2$ induces a spin-independent checkerboard pattern of electric current. This explains the nomenclature ``s-DDW'', i.e., ``singlet DDW''. In the absence of $\vec{m}_1$ the energy spectrum  is nodal, with the nodes centered along the Brillouin zone diagonals. In Ref.\cite{Laughlin} this feature is regarded as the signature of pseudogap. The order parameter $\vec{m}_1$  is absent in Ref.\cite{Laughlin}. It describes a spin-dependent  second neighbor hopping. After fixing the direction of $\hat{m}_1$, the hopping amplitude has opposite sign in the (1,1)/(1,-1) directions and modulates with momentum $(\pi,\pi)$. In addition, the hopping amplitudes change sign when electron's spin polarization along $\hat{m}_1$ reverses.
We schematically represent $H_{\rm s-DDW}$ in \Fig{fig1}(a).
A spatially uniform $\vec{m}_1$ opens a gap in the energy spectrum.  Moreover, as long as $|\vec{m}_1|$ is small compared with $m_2$, doping will create Fermi pockets around the nodes. 
\\

The above model should be viewed as the mean-field theory of certain interacting Hamiltonian similar to that discussed in Ref.\cite{Laughlin2}. The order parameter $\vec{m}_1$ breaks
the translation and  4-fold rotation  symmetries of the lattice.
Moreover, it also breaks the SU(2) spin rotation symmetry down to U(1), namely, rotation around the $\hat{m}_1$ axis. The order parameter $m_2$ breaks the translation and time reversal symmetry. However, $H_{\rm s-DDW}$ respects the combined operation of time reversal and translation. \\

\subsubsection{The edge states}
In \Fig{fig1}(b) we plot the energy spectrum of \Eq{model} with $m_1=0.15\hat{z}$ and $m_2=0.5$ in the cylindrical geometry, namely, open boundary condition along $\hat{y}$ and periodic boundary condition along $\hat{x}$. Here $\hat{x}$ and $\hat{y}$
are 45 degrees rotated from the principal axes of the square lattice, and $k_x$ is a momentum in the antiferromagnetic Brillouin zone. There is a pair of counter-propagating helical edge modes localized on each of the two edges, as shown in \Fig{fig1}(b). They are reminiscent of the edge modes in a quantum spin Hall insulator. These edge modes are protected from back scattering by the residual U(1) spin rotation symmetry, hence the system is a topological insulator.
In \Fig{fig1}(c) we plot the energy spectrum in the presence of a $z$-diection magnetic field. Clearly, the edge modes are Zeeman split. \\

Like the quantum spin Hall insulator, the electric Hall conductance of model 1 is zero. However, due to the Zeeman splitting, a magnetic field induces a non-zero current on each edge. This is because the spin up and spin down edge electron density are no longer equal. However, in the cylindrical geometry, this magnetic-field-induced edge current cancels among the two edges.  In the disk geometry, the magnetic-field-induced edge current circulates around the perimeter, as shown in \Fig{fig1}(d). This edge current implies the presence of a bulk orbital magnetization.\\

\begin{figure}[t]
\includegraphics[height=1.1in]{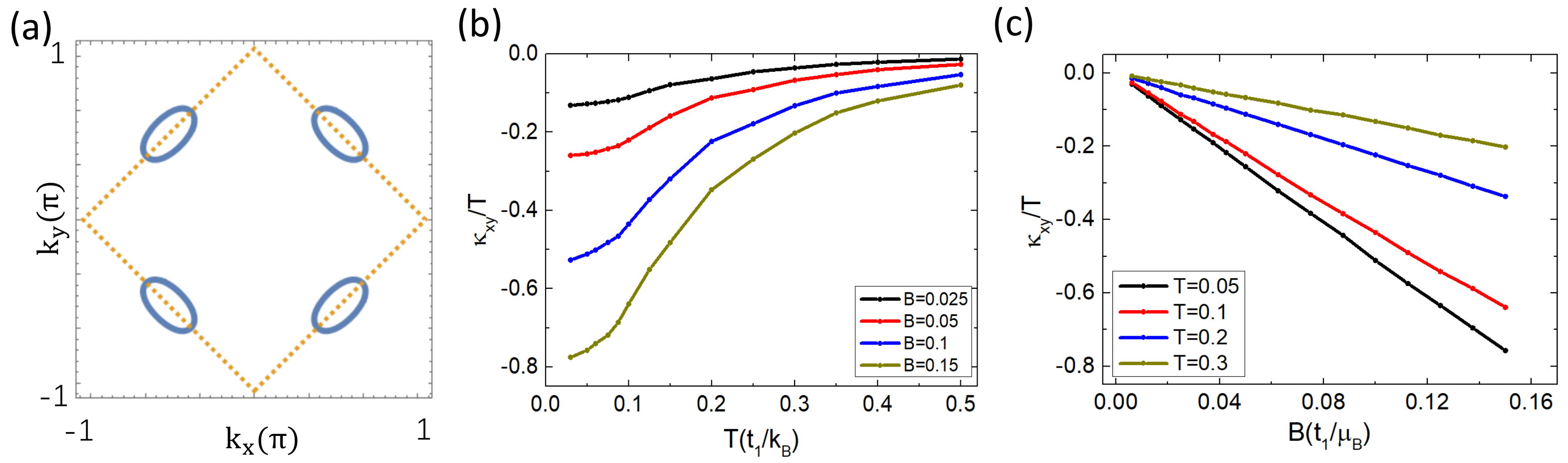}
\caption{ (a) The Fermi surface of the Hamiltonians  in \Eq{model} and \Eq{model2} for doping level $p=0.06$. The parameters used are $\vec{m}_1 = 0.15 \hat{z}, m_2 = 0.5$ for \Eq{model} and $m_1 = 0.15, \vec{m}_2 = 0.5\hat{z}$ for \Eq{model2}.  The blue solid line represents the Fermi surface and the yellow dashed line encloses the antiferromagnetic Brillouin zone. (b) The thermal Hall conductivity $\kappa_{xy}/T$ (in units of $k_B^2/\hbar$) as a function of temperature $T$ at several magnetic fields $B$ for doping level $p=0.06$. (c)  $\kappa_{xy}/T$ (in units of $k_B^2/\hbar$) as a function of applied magnetic field $B$ at several temperatures for doping level $p=0.06$. }
\label{fig2}
\end{figure}

\subsubsection{The thermal Hall effect and the Fermi pockets}

Following Ref.\cite{Lee-2019} we show that upon doping \Eq{model} exhibits an unusual thermal Hall effect. Doping is achieved by adding a chemical potential term to $H_0$, namely $-\mu\sum_{i}C^+_{i\alpha}C_{i\alpha}$. In the first version of the manuscript we attribute the thermal Hall effect to the edge thermal conduction. This leads to the conclusion that the thermal conductivity is non-zero even in the insulating state. The authors of Ref.\cite{Lee-2019} pointed out to us that the thermal conduction due to the helical edge states should be negligible for weak fields. This is because despite the Zeeman shift, the energy current due to the particle-hole excitations near the chemical potential are the same for both spins (due to the cancellation between the density of states and the Fermi velocity in 1D). Thus the spin up and spin down electron's contributions to the thermal conductivity cancel. However, when the chemical potential lies within the bulk bands, and when the Berry curvature is non-zero in the energy range of [$E_f-B,E_f+B$] the bulk thermal Hall conductivity is non-zero. However, this bulk contribution requires finite doping. \\

It can be shown straightforwardly that for \Eq{model} with $(\vec{m}_1,m_2)=(m_1\hat{z},m_2)$ the energy dispersion and the Berry curvature $B_{n\alpha}(\v k)$ are given by
\be
&&E_{n\alpha}(\v k)= -2t_2 (c_{x+y}+c_{x-y})-\mu - 2 n  R_{x,y}-\alpha\mu_B B\nn
&&B_{n\alpha}(\v k) =
n\alpha\left(2m_2 m_1 t_1(s^2_x + s^2_y -s^2_x s^2_y\right)/R_{x,y}^3\nn&&R_{x,y}=\sqrt{t_1^2(c_x+c_y)^2+m_2^2(c_x-c_y)^2+4m_1^2s_x^2s_y^2}
\label{disp}\ee
Here, $n=\pm 1$ refers to the lower and upper band, and $\alpha=\pm 1$ are the spin polarization along  the $\hat{z}$ (i.e., $\hat{m}_1$) direction. In addition, we have used the abbreviations $c_{x(y)}=\cos k_{x(y)}$, $s_{x(y)}=\sin k_{x(y)}$,$c_{x\pm y} = \cos (k_x \pm k_y)$. In terms of $E_{n\alpha}(\v k)$ and $B_{n\alpha}(\v k)$ the thermal Hall conductivity (in units of $k_B^2/\hbar$) is given by\cite{Niu}:
\bea
&&\frac{\kappa_{xy}}{T}=\frac{1}{4T^3}\int d\epsilon \frac{(\epsilon-\mu)^2}{\cosh^2[\beta(\epsilon-\mu)/2]} (\sigma_{xy\uparrow}(\epsilon)+\sigma_{xy\downarrow}(\epsilon))\nn&&\sigma_{xy\alpha}(\epsilon) = -\sum_{nk} B_{n\alpha}(\v k)\theta(\epsilon - E_{n\alpha}(\v k)).
\label{kappa}
\eea

In the following we adjust the chemical potential $\mu$ so that the doping level is  $p=0.06$.
In \Fig{fig2}(a) we show the Fermi surface for this doping level. It consists of hole pockets centered along the Brillouin zone diagonals. In \Fig{fig2}(b) we show $\kappa_{xy}/T$ as a function of temperature at several magnetic field values.  First, the sign of $\kappa_{xy}$ is negative. Second,  at a fixed magnetic field $|\kappa_{xy}|/T $ increases with decreasing temperature.  In \Fig{fig2}(c) we show the dependence of $\kappa_{xy}/T$ as a function of magnetic field at different temperatures. The result monotonic increases with $B$. Features (a)-(c) are consistent with what's seen in Ref.\cite{Taillefer-2019}. \\

A more stringent test of the theory  is the actual size of the predicted $\kappa_{xy}/T$. According to Fig. 1(b) of Ref.\cite{Taillefer-2019}, under a 15T magnetic field the  $|\kappa_{xy}/T|$ at the lowest measurement temperature is about 0.7 $k_B^2/\hbar$ per copper-oxide plane. If we set $t_1\sim$ 200 meV, 15T corresponds to $B=0.0075 t_1/\mu_B$ and $T_{\rm min}=14K$ corresponds to $T= 0.007 t_1/k_B$. We have checked that for these parameters the largest $|\kappa_{xy}/T|$ obtainable by varying $|\vec{m}_1|$  at a fixed $m_2=0.5$ is 0.1 $k_B^2/\hbar$. \\

Thus \Eq{model} has the potential to explain the following two very unusual experimental features observed in the underdoped regime of the cuprates where there is no charge order. (1) Hole pockets centered along the Brillouin zone diagonals with area equal to the doping concentration. 
 (2) The anomalous thermal Hall effect observed in Ref.\cite{Taillefer-2019}. \\

In addition, \Eq{model} also predicts the existence of a checkerboard pattern of staggered orbital magnetic moments. These moments have been experimentally searched for, but so far there is no convincing  evidence for it. For this reason we proceed to consider the ``tripet-DDW'' model in the following section.

\section{Model 2\cite{Hsu}: a modified triplet-DDW}
The  model introduced in Ref.\cite{Hsu} is given by
 \bea
&&H = H_0 + H_{\textrm{t-DDW}} \nonumber\\
&&H_0 = -t_1 \sum_{\avg{ij}} C^\dagger_{i\alpha} C_{j\alpha} -t_2 \sum_{\avg{\avg{ij}}} C^\dagger_{i\alpha} C_{j\alpha} + h.c. \nonumber\\
&&H_{\textrm{t-DDW}} =\sum_i(-1)^{i_x+i_y}\Big\{(i~\vec{m}_2\cdot\vec{\s}_{\alpha\beta})\Big[C^\dagger_{i\alpha} C_{i+x\beta} \nn&&-C^\dagger_{i\alpha}C_{i+y\beta}\Big]+m_1\Big[C^\dagger_{i\alpha}C_{i+x+y\alpha} - C^\dagger_{i\alpha} C_{i-x+y\alpha}\Big] + h.c.\Big\} \nonumber\\
\label{model2}
\eea

Here the term proportional to $i \vec{m}_2$ is a spin-dependent DDW order parameter (hence the nomenclature of ``t-DDW'', i.e., ``triplet DDW'').
The important difference with the model in \Eq{model} is the cancellation of the orbital magnetic moments because the pattern of circulating current is opposite for spin up and spin down electrons. Thus it removes the unwanted feature of a predicted, but unobserved, orbital magnetic moment.
  The order parameter proportional to $m_1$  is a spin-independent second neighbor hopping. It also opens an energy gap at the nodes.\\

\begin{figure}[t]
\includegraphics[scale=0.28]{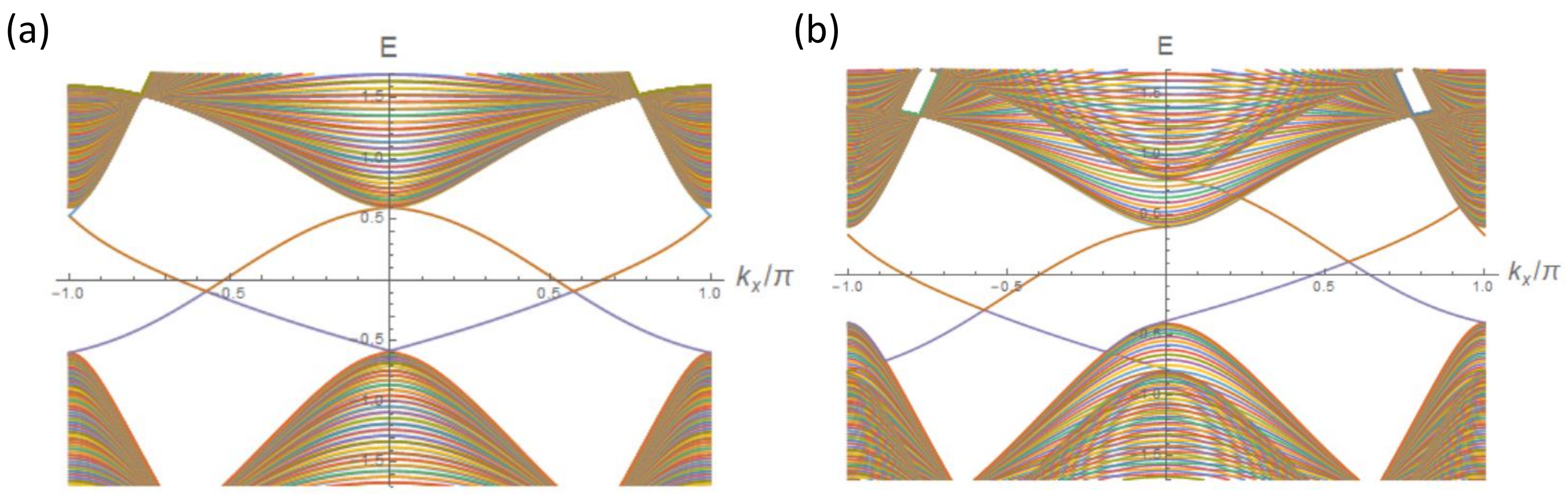}
\caption{ (a) The projected band structure of the Hamiltonian in \Eq{model2} with periodic boundary condition in x-direction and open boundary condition in y-direction. The number of rows in y-direction is $n_y=100$. The parameters used are  $m_1 = 0.15, \vec{m}_2 = 0.5 \hat{z}$. The edge branches for $k_x>0$ and $k_x<0$ are associated with the spin up and spin down electrons, respectively. For each spin direction the counter propagating branches are localized on different edges. (b) The projected band structure of \Eq{model2} in the presence of a magnetic field in the $\hat{z}$ direction. The associated Zeeman energy is set to $0.2$. }
\label{fig3}
\end{figure}

The term proportional to $m_1$ breaks
the translation, 4-fold rotation and mirror symmetries along the x and y axes.
 In addition, the order parameter $i\vec{m}_2$ breaks the SU(2) spin rotation symmetry down to U(1). However, interestingly, $i\vec{m}_2$ preserves the time-reversal symmetry. This last statement explains why $\vec{m}_2$ does not generate any orbital magnetic moment. It is also the reason why $\vec{m}_2$ is not visible to experimental probes such as neutron scattering and NMR. \\
\\

\subsubsection{The edge states}

In \Fig{fig3}(a) we plot the energy spectrum of \Eq{model2} with  $m_1=0.15$ and $\vec{m}_2=0.5\hat{z}$ in the cylindrical geometry.
Again, on each edge there is a pair of counter-propagating helical edge modes. These edge modes are protected against back scattering by the time reversal and/or the residual spin U(1) symmetries. In the absence of disorder it is also prevented from back scattering because the Fermi momenta of the right and left movers are different.  In \Fig{fig3}(b) we plot the energy spectrum in the presence of a magnetic field. Here we have assumed $\hat{m}_2$ to lie in the magnetic field direction, namely, $\hat{z}$.  Clearly, the edge modes are Zeeman split.
Like model 1, this topological insulator shows zero electric Hall conductance. In the disk geometry there is a magnetic-field-induced circulating boundary current, which reflects the existence of a non-zero bulk orbital magnetization.\\

\subsubsection{The thermal Hall effect and the Fermi pockets}

It turns out that for \Eq{model2}, with $(m_1,\vec{m}_2)=(m_1,m_2\hat{z})$, the
band dispersion and the Berry curvature are exactly the same as in those for \Eq{model} with $(\vec{m}_1,m_2)=(m_1\hat{z},m_2)$. Therefore at the same doping level ($p=0.06$)
and with the same $m_1$ (0.15) and $m_2$ (0.5), the Fermi surface and $\kappa_{xy}/T$ are identical to those shown in \Fig{fig2}. However, model 2 does not possess the staggered orbital magnetic moment.\\

\section{The pinning of $\vec{m}_1$ and $\vec{m}_2$ by the magnetic field}
The vector order parameter $\vec{m}_1$ in \Eq{model} and $\vec{m}_2$ in \Eq{model2} are free to rotate without causing any energy. This implies the presence of Goldstone modes. In the presence of these soft modes one needs to worry about the disordering of these vector order parameters at non-zero temperatures (particularly in two spatial dimensions).\\

To address these issues, we focus on zero doping. The generalization to the doped case is straightforward.
In the following we shall focus on \Eq{model2}. To obtain the corresponding statements for \Eq{model}  one just need to exchange the roles of $m_1$ and $m_2$.\\

As discussed earlier, a non-zero magnetic field induces a bulk orbital magnetization. The latter is given by\cite{vanderbilt}
\be
{\cal M}=-\sum_{\alpha=\hat{m}_2\cdot\vec{\s}=\pm 1}{e\over h c}C_{\alpha} \Delta E_{Z\alpha}.\label{M}\ee
Here $c$ is the speed of light, $h$ is the Planck constant, $e$ is the electron charge, and $C_{\alpha}$ is the Chern number of the spin $\alpha$ band. In addition, the Zeeman energy,  $ \Delta E_{Z\alpha}$, is given by $-\alpha\mu_B\hat{m}_2\cdot\vec{B}$ where $\mu_B$ is the effective electron magnetic moment, and $C_{-1}=-C_{+1}$.
Since the reversal of the sign of $\alpha$ causes both $C_\alpha$ and $ \Delta E_{Z\alpha}$ to change sign, \Eq{M} can be simplified to
\be
{\cal M}=-{2e\over  h c}C_{+1} \Delta E_{Z+1}={2e\mu_B\over  h c}C_{+1}\hat{m}_2\cdot\vec{B}.\ee
Importantly, the sign of $C_{+1}$ is determined by that of $m_1$, namely,
\be
C_{+1}={m_1\over |m_1|}.\ee
Putting these results together we have
\be {\cal M}={2 e\mu_B\over  h c}{m_1\over |m_1|}\hat{m}_2\cdot\vec{B}.\ee
The above orbital magnetization interacts with the magnetic field via the Zeeman coupling to yield the following energy density
\be
\Delta {\cal E}_{\rm Zeeman}=-{\mu_B B^2\over  \pi c}{m_1\over |m_1|}(\hat{m}_2\cdot\hat{B}).\label{int}\ee
\Eq{int} implies that in the presence of a magnetic field it is energetically favorable for $m_1\hat{m}_2$ to point in the same direction as $\hat{B}$.  This eliminates the Goldstone modes and fixes the sign of $\kappa_{xy}$. Thus the sign of $\kappa_{xy}$ should not be random among different cool downs.\\

In two space dimensions the SO(3) symmetry breaking in both \Eq{model} and \Eq{model2} are only present in a non-zero applied magnetic field. This provides examples where the zero field and finite field electronic states can be different.
 In zero magnetic field it is interesting to study the fate of the topological insulators when $\vec{m}_1$ or $\vec{m}_2$ is thermally disordered. This study reveals an important  difference between model 1 and model 2. For model 1 the residual U(1) spin symmetry is broken by any disordered configuration of $\vec{m}_1$.  Hence we expect the edge states to loose symmetry protection. In contrast,  for model 2  the edge states stay protected (by the time reversal symmetry) even when the U(1) spin rotation symmetry is lost. This difference is confirmed by examining the thermal-averaged edge spectral function of model 1 and model 2 in the cylindrical geometry, namely,
   \be
&&\overline{A}(k_x,\omega) = \frac{\sum_{\{\vec{m}_{a,i}\}}W[\{\vec{m}_{a,i}\}] A(k_x,\omega)_{\{\vec{m}_{a,i}\}}}{\sum_{\{\vec{m}_{a,i}\}} W[\{\vec{m}_{a,i}\}]}\nn&&W[\{\vec{m}_{a,i}\}] = e^{-\beta J \sum_{\avg{ij}} \vec{m}_{a,i}\cdot \vec{m}_{a,j}}.
\label{thermal}
\ee
Here $\{\vec{m}_{a,i}\}$, with $a=1 {\rm ~or~} 2$, are the spatial configurations of the vector order parameter in \Eq{model} or \Eq{model2}, and  $W[\{\vec{m}_{a,i}\}]$ is the Boltzmann weight. $A(k_x,\omega)_{\{\vec{m}_{a,i}\}}$ is the spectral function under a fixed configuration of $\{\vec{m}_{a,i}\}$. \\

\begin{figure}[t]
\includegraphics[height=1.75in]{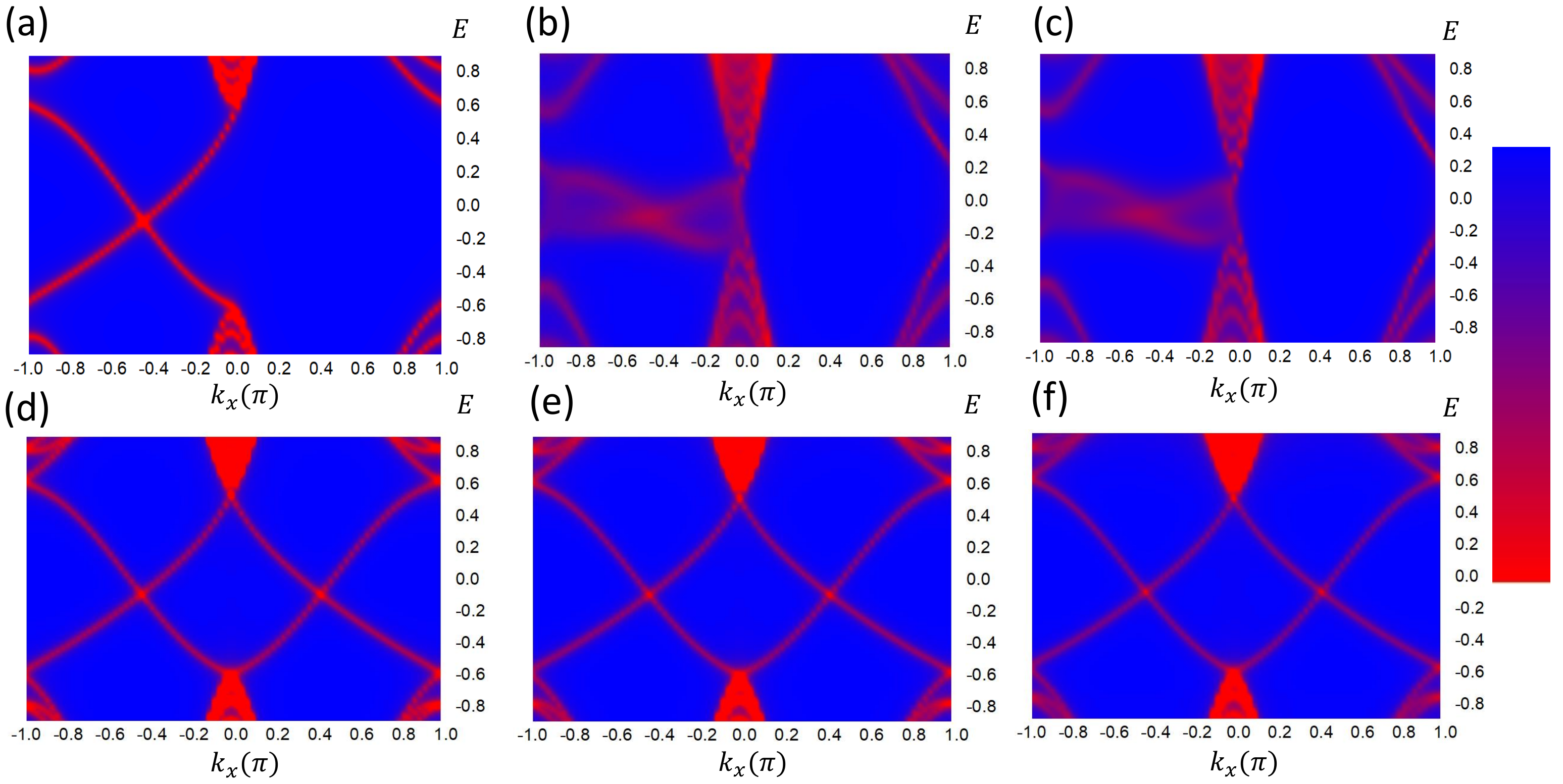}
\caption{ The thermal averaged electron spectral function in models \Eq{model} and \Eq{model2} in a finite cylinder (open boundary condition in $\hat{y}$ and periodic boundary condition in $\hat{x}$). The parameters used are $\vec{m}_1 = 0.15 \hat{z}, m_2 = 0.5$ for \Eq{model} and $m_1 = 0.15, \vec{m}_2 = 0.5\hat{z}$ for \Eq{model2}. The linear dimension of the cylinder is  is $n_x = 80$ and $n_y = 20$. The ensembles of $\{\vec{m}_{1,i}\}$ and $\{\vec{m}_{2,i}\}$ are generated with the Boltzmann weight given in \Eq{thermal}. Panels (a)-(c) are the results for \Eq{model} while panels (d)-(f) are for \Eq{model2}. The inverse temperatures used in the calculations are $\beta J = \infty$ (zero temperature) in panels (a) and (d), $\beta J = 16$ and in panels (b) and (e), and $\beta J = 8$ in panels (c) and (f). The total number of sampled configurations is 30000.}
\label{fig4}
\end{figure}

Our calculation is performed after fixing the amplitude $|\vec{m}_1|$ or $|\vec{m}_2|$. We sample the directions of $\hat{m}_{1}$ or $\hat{m}_{2}$ according to the Boltzmann weight by the Metropolis algorithm, and the number of sampled configurations is 30000. As shown in \Fig{fig4}(b,c), the edge modes in \Eq{model} are disorder scattered at non-zero temperatures. In contrast, the edge modes in \Eq{model2} remain  sharp as shown in \Fig{fig4}(e,f). We attribute this difference to the fact that for \Eq{model2} thermal disordering of $\vec{m}_2$ does not jeopardize one of the protection symmetry, namely, the time reversal symmetry.
\\

\section{\bf The Neel ordered phase}

\begin{figure}[t]
\includegraphics[height=1.1in]{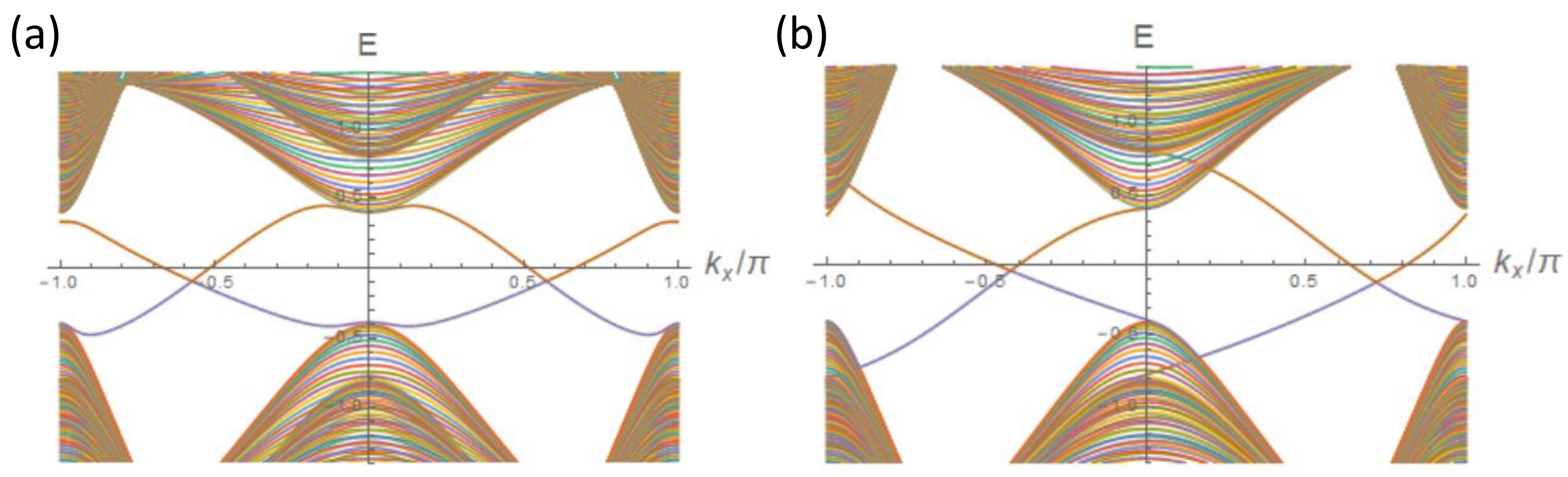}
\caption{ The projected band structure of $H_{\rm t-DDW}$ in the presence of Neel order. The number of rows in the open ($\hat{y}$) direction is $n_y=100$. In these plots $\vec{m}_s$ is set to $ 0.2\hat{x}$ and $m_1=0.15$. In panel (a) $\vec{m}_2=0.5\hat{z}$ and in panel (b) $\vec{m}_2=0.5\hat{x}$. }
\label{fig5}
\end{figure}

The topological nature of the model 1 and model 2  survives the presence of the Neel long range order,
\be
H_{\textrm{Neel}} = \sum_i (-1)^{i_x+i_y} \vec{m}_s\cdot\vec{\s}_{\alpha\beta}C^\dagger_{i\alpha}C_{i\beta},\ee
as long as $\vec{m}_s$ is not too strong.
For example, in \Fig{fig5}(a) and (b) we show the edge modes dispersion of model 2 in the presence of a non-zero $\vec{m}_s=0.2 \hat{x}$. The parameters used are $m_1 = 0.15$ and $\vec{m}_2=0.5\hat{z}$ in panel (a) and $m_1 = 0.15$ and $\vec{m}_2=0.5\hat{x}$ in panel (b) .  \\

Despite the persistence of the edge states, our models predict the absence of thermal Hall effect in the undoped limit, agreeing with the result of Ref.\cite{Lee-2019}.  This is because when the sample is undoped, the chemical potential lies in the gap of the  Zeeman shifted spin up and spin down spectrum (at least when the Zeeman energy is small compared to the gap energy). Under such condition \Eq{kappa} predicts zero thermal Hall conductance because the $\epsilon$-integrals for spin up and spin down electrons yield values with opposite sign but the same (quantized) magnitude, hence they cancel\cite{Lee-2019}.

\section{The effect of residual electronic correlation on the edge states}

The main effect of the electronic correlation is to render the system in the mean-field state described by \Eq{model} or \Eq{model2}. In the following we discuss the effects of residual electronic correlation on the edge dynamics. The fact that this is necessary is because the edge modes are gapless. \\

The edge Hamiltonian is given by
\bea
H_E = iv_F \int dx~\Big[\psi^\dagger_{L\downarrow}(x) \partial_x \psi_{L\downarrow}(x) -  \psi^\dagger_{R\uparrow}(x) \partial_x \psi_{R\uparrow}(x)\Big]\nonumber\\
\label{freefermion}
\eea
where $\psi_{L\downarrow}$ and $\psi_{R\uparrow}$ are the annihilation operators of the left (spin down) and right (spin up) moving edge electrons, and  $v_F$ is the mean-field edge velocity. Due to the time reversal and/or the residual spin U(1)  rotation symmetry, the single-particle backscattering terms, $\psi^\dagger_{R\uparrow} \psi_{L\downarrow} + \psi^\dagger_{R\uparrow} \psi_{L\downarrow} $, and $i \psi^\dagger_{R\uparrow} \psi_{L\downarrow} - i\psi^\dagger_{R\uparrow} \psi_{L\downarrow}$, are not allowed.\\

The most relevant, symmetry-allowed, four fermions interactions is given by
\bea
&&H_{\textrm{int}} = \int dx~\Big\{~g_2 \psi^\dagger_{R\uparrow}(x)\psi_{R\uparrow}(x) \psi^\dagger_{L\downarrow}(x) \psi_{L\downarrow}(x) \Big] \Big\}.
\eea
It renormalizes the edge velocity and the Luttinger liquid parameter: \be &&v=v_F\sqrt{(1-\frac{g_2}{v_F})(1+\frac{g_2}{v_F})}\nn&& K = \sqrt{\frac{v_F-g_2}{v_F+g_2}}.\ee
\\ 

The usual process that opens the charge gap is the umklamp scattering $g_u\psi^\dagger_R \psi^\dagger_R \psi_L \psi_L$. It is forbidden, due to the Fermi statistics, in the present situation due to the spin-momentum locking of the edge electrons. Hence residual correlation does not affect the edge states qualitatively.
\\

\noindent{{\bf Final discussions}\\

The topological insulators described by \Eq{model} and \Eq{model2} have the following attractive features. (1) Under low level of p-type doping they predict  hole pockets centered along the Brillouin zone diagonals. This is consistent with the Hall coefficient measurement\cite{Taillefer2} which shows a carrier density $p$ rather than $1+p$ in the doping range where the anomalous thermal Hall effect is observed.
(2) These models can explain the anomalous thermal Hall effect in all samples {\it except the undoped La$_2$CuO$_4$}.
It is also important to point out we did not provide any microscopic justification for the models in  \Eq{model} and \Eq{model2}. Whether there exists, e.g., one-band or three-band Hubbard-like models which realize \Eq{model} or \Eq{model2} as the stable mean-field solution is unclear to us at present.\\

Finally, we take note of several related experimental facts. (1) There is a report from thermal transport that the pseudogap temperature $T^*$ coincides with the onset of 90 degree rotation symmetry breaking\cite{Taillefer2}. Could this be due to the symmetry breaking induced by $\vec{m}_1$ in model 1 or $m_1$ in model 2 ?  Ref.\cite{Hsieh} reports that in the pseudogap regime, YBCO exhibits inversion symmetry breaking below $T^*$. In addition, the polar Kerr effect suggests the breaking of time reversal symmetry\cite{Kerr}.
Although  model 1 breaks time reversal symmetry, it does not break inversion. Model 2 does not break time reversal nor inversion. Although it is possible to add inversion and time reversal breaking features to the two models (for example by making $m_1$ complex) we prefer not to do so for the sake of simplicity.
Lastly, in an ARPES experiment on Bi2201 a small nodal gap is observed in the doping range close to the AFM phase boundary\cite{Zhou-2013}. Could it be the gap caused by $\vec{m}_1$ (or $m_1$)?  \\

Before the end, we take note of three recent interesting theory papers\cite{Sachdev-2018,Lee-2019,Xu-2019} on the same subject.  Our theory, in particular model 1,  bears a strong resemblance to that in Ref.\cite{Lee-2019}.
Our explanation of the thermal Hall conductance is the same as theirs. However there is an important difference between our theory and Ref.\cite{Lee-2019}, namely, the fermions in our theory are the physical electrons.\\

\noindent{{\bf Acknowledgement}} \\

We are in debt to Prof. Steve Kivelson for bringing Ref.\cite{Taillefer-2019} to our attention. In addition, he  pointed out two references which eventually lead us to Ref.\cite{Hsu}. We thank Prof. Bob Laughlin for enlightening discussions. He raised the important question concerning the sign of $\kappa_{xy}$ upon different cool down, and told us about the possible existence of Ref.\cite{Hsu}. We thank  Prof. Chandra Varma for enlightening discussions including the question on the meaning of the sign of $\kappa_{xy}$. Finally, we are very grateful to the authors of Ref.\cite{Lee-2019} for pointing out that the thermal conduction due to helical edge states should vanish.  This work was primarily funded by the U.S. Department of Energy, Office of Science, Office of Basic Energy Sciences, Materials Sciences and Engineering Division under Contract No. DE-AC02-05-CH11231 (the Quantum Materials  program). We also acknowledge support from the Gordon and Betty Moore Foundation's EPIC initiative, Grant GBMF4545.

\end{document}